\documentstyle[epsfig,12pt]{article}

\newfont{\eurm}{eurm10}

\newcommand{\ms}{m$\;$s$^{-1}$}
\newcommand{\mss}{m$\;$s$^{-2}$ }
\newcommand{\apj}{{\it Astrophys.~J.} }

\newcommand{\solphys}{{\it Sol. Phys.} }
\newcommand{\nat}{{\it Nature} }
\newcommand{\mnras}{{\it Mon. Not. R. Astron. Soc.} }

\newcommand{\sci}{{\it Science} }
\newcommand{\etal}{{\it et al.} }

\date{}

\begin{document}

\noindent
{\huge {\sf \bf \sf Solar supergranulation\\ as a wavelike phenomenon} }
\vskip 5mm
\noindent
{\normalsize {\bf L. Gizon$^\ast$, T.~L. Duvall Jr.$^\dag$, and J. Schou$^\ast$} }
\vskip 2mm
\noindent {\footnotesize \it
$^\ast$W.~W. Hansen Experimental Physics Laboratory, Stanford University, Stanford, CA 94305, USA \\
$^\dag$Laboratory for Astronomy and Solar Physics,  NASA Goddard Space Flight Center, Greenbelt, MD 20771, USA 
}



\vskip 5mm
{\bf \noindent
Supergranulation on the surface of the Sun is an organized cellular flow pattern with a characteristic scale of 30~Mm\cite{leighton1}. It is superficially similar to the well understood granulation\cite{stein} that operates at the 1.5~Mm natural scale of convection, which has led to the conventional view that supergranulation has its origin in the convective motion of cells of gas\cite{leighton2,borght}, though this does not explain the observation that the supergranulation pattern appears to move faster around the Sun than the bulk of its surface\cite{duvall80,snodgrass, beck_schou}. A wave origin has been proposed\cite{wolff} for supergranulation that may explain the superrotation, but it has never had much support. Here we report that the supergranulation pattern 
has  oscillatory components with periods of 5-10 days, for which the best explanation is a spectrum of traveling waves. We show that there is excess power in the prograde and equatorward directions, which explains the observation of superrotation.}
\vspace{2 mm}

\noindent
To study supergranulation, we use a 60-day sequence of line-of-sight
Doppler velocity images of the Sun's surface, obtained by the 
Michelson Doppler Imager\cite{scherrer} on board the Solar and 
Heliospheric Observatory (SOHO).
We apply the technique of time-distance helioseismology\cite{duvall93} 
to obtain maps of the horizontal divergence of the flows just below 
the solar surface.
Unlike raw Doppler images, the divergence signal has uniform sensitivity 
across the solar disk and has few systematic problems.
Every 6~hr, an 8~hr section of data is used to obtain a divergence map. This
is done by measuring the time it takes for surface gravity waves (f modes) to propagate horizontally between a given point on the solar surface and a concentric annulus with radius 8~Mm. The difference in travel times between inward and outward propagating waves is a proxy for the horizontal divergence of the flow field in the first 2~Mm beneath the surface. The details of this analysis are described elsewhere\cite{duvall_gizon}.  
Supergranules appear as cellular patterns of horizontal outward flow in the divergence maps.

We considered five latitude bands, 90$^\circ$ in longitude by $10^\circ$ in latitude, centered at latitudes $0^\circ$, $\pm15^\circ$, and $\pm30^\circ$.  For each latitude band, the divergence signal is Fourier transformed in three dimensions to make power spectra as a function of angular frequency, $\omega$, and horizontal wavevector, ${\bf k}$. Figure~1 shows a cylindrical cut in a power spectrum at a constant wavenumber, $k=\|{\bf k}\|$, typical of the supergranulation.  The power peaks at a non zero frequency implying that the supergranular divergence field oscillates from positive to negative values.  This observation alone argues against a simple model of convection\cite{harvey,kuhn00}.  Figure~1 is characteristic of waves propagating in all directions at a well defined frequency. The variation of power with the direction of propagation, $\phi$, indicates a strong anisotropy. We thus suggest that we are observing a superposition of traveling waves, and show below that this hypothesis provides a solution to the superrotation puzzle.

Waves that propagate horizontally through a moving fluid have a dispersion relation of the form $\omega=\omega_0(k) + {\bf k} \cdot {\bf v}$, where $\omega_0$ is the wave frequency in the co-moving frame of reference and ${\bf v}$ is the advection velocity. At fixed wavenumber, we thus interpret the sinusoidal variation of the frequency with $\phi$ to be due to a horizontal flow (Fig.~1). This technique for measuring material flows is equivalent to the ring-diagram technique\cite{schou_bogart} used in helioseismology. Figure~2 shows the inferred rotation and meridional circulation, both similar to that of the small magnetic features\cite{komm_rot,komm_mer}. The rotation corresponds to the helioseismic value at a depth of 20~Mm, suggesting that the waves probe the region of increased rotation just beneath the solar surface. 
Correlation tracking of the supergranulation pattern, on the other hand, gives a larger rotation rate, larger that the rotation of the plasma measured at any depth in the solar interior\cite{beck_schou}, as well as a  puzzling equatorward meridional motion (Fig.~2). 

The anomalous results from correlation tracking can be understood as a consequence of the pronounced anisotropy of wave power (Fig. 3). The flow velocities inferred from the correlation tracking are effectively a power-weighted average of the true advection velocity, {\bf v}, and the wave propagation speed (about 65~\ms).  The supergranulation pattern appears to superrotate because wave power in the prograde direction is almost twice as large as in the retrograde direction.   Since the change in the rotation over the outer 50~Mm is a significant fraction of the phase speed, a buried source of excitation could introduce an anisotropy. Note also that a model of standing-wave oscillatory convection advected by a flow would fail to reproduce the anisotropy in wave power.

We have shown that the power spectrum of the divergence field of the supergranultion is consistent with an anisotropic spectrum of traveling waves.
One possibility\cite{wolff} for the nature of the waves is that they are driven by the Coriolis force, analogous to the oceanic Rossby waves\cite{chelton} and perhaps related to the recently observed solar r modes\cite{kuhn00}.
However, wave periods are found to be significantly smaller than the rotation period  (Fig.~4) and we found no evidence for a latitudinal variation in the dispersion relation. 
Another possibility is traveling-wave oscillatory convection\cite{matthews}.
For example, numerical simulations of turbulent convection in oblique magnetic fields \cite{hurlburt} exhibit solutions that take the form of traveling waves, where the tilt of the convection cells, their wave speed, and direction depend on the field strength and obliquity. 
Finally, we note that the anisotropy in wave power is important as it could provide a mechanism of angular momentum transport in the upper convection zone, that could help explain the variation of rotation with latitude\cite{gilman}.

\footnotesize
\scriptsize

\normalsize

\footnotesize
\vspace{5mm}
{\normalsize \bf Acknowledgments} \newline 
We thank J.~G. Beck, P. Milford, P.~H. Scherrer, C.~J. Schrijver, and N.~O. Weiss for useful discussions.  SOHO is a mission of international cooperation between the European Space Agency and NASA.  MDI is supported by the Office of Space Sciences of NASA.

\vspace{5mm}
$\!\!\!\!\!\!\!\!$
$\!\!\!\!\!\!\!$
{\normalsize \bf Competing interests statement}  \newline
The authors declare that they have no competing financial interests.

\vspace{5mm}
$\!\!\!\!\!\!\!\!$
$\!\!\!\!\!\!\!$
Correspondence and requests for material should be addressed to L.G.
\newline (e-mail: lgizon@solar.stanford.edu).

\newpage

\begin{figure}[t]
\centerline{\epsfig{file=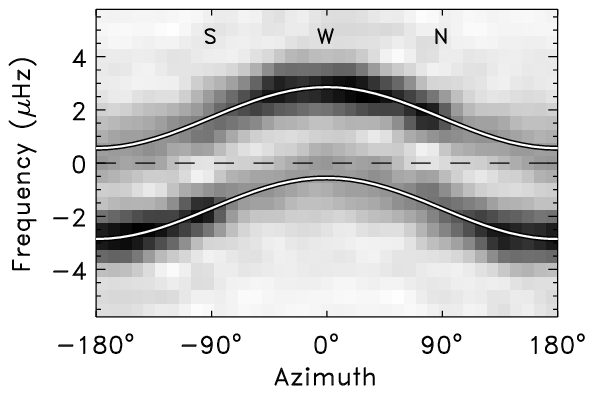, width=15.cm}}
\vspace{1.cm}
{\normalsize {\bf Figure 1} Cylindrical cut at constant wavenumber in the 3D power spectrum of the divergence signal near the equator, with $k=120/R$ where $R$ is the solar radius.  Black indicates larger power. The azimuth, $\phi$, is the angle between ${\bf k}$ and the direction of solar rotation (West). 
 The top ridge is the power in the waves traveling in
the direction given by $\phi$. The bottom ridge is for waves
traveling in the opposite direction and is identical to the upper
ridge, but shifted by $180^\circ$. 
We applied an MTF correction estimated from the data.
To measure the flows, we first fit, at each $\phi$, the sum of two Gaussian functions in frequency with independent central frequencies and common line widths. We then fit curves (double lines) of the form $\omega=\pm\omega_0 + k v_x \cos\phi  + k v_y \sin\phi$ to the central frequencies, where $\omega_0$ is the wave frequency, $v_x$ is the velocity in the direction of rotation, and $v_y$ is the meridional circulation.   For each latitude, $v_x$ and $v_y$ are measured to be roughly independent of $k$ in the range  $50 < kR < 200$, consistent with the interpretation as a flow.  The raw data was tracked at the Carrington rotation rate, leading to a small prograde frequency shift for the equatorial band.
}
\end{figure}

\newpage

\begin{figure}[t]
\centerline{\epsfig{file=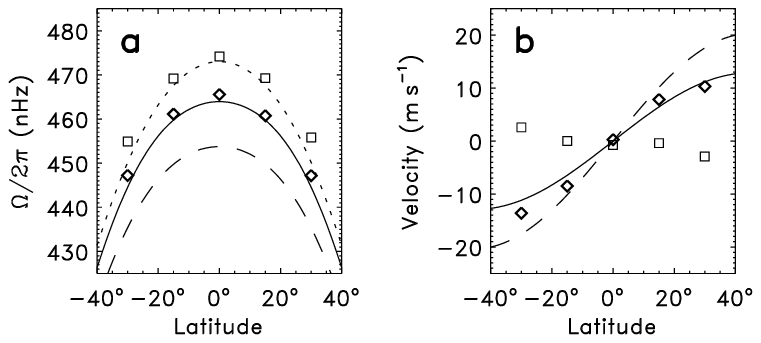, width=15.cm}}
\vspace{1.cm}
{\normalsize {\bf Figure 2} 
Flows inferred from the advection of the supergranulation waves.
{\bf a}, Solar rotation (diamonds). 
The magnetic feature rate\cite{komm_rot} (solid line) 
is very similar and the photospheric rate\cite{snodgrass} (dashed) 
is lower.  The squares show the results using 
correlation tracking of the supergranulation pattern with a 24~hr delay, 
consistent with an earlier estimate\cite{snodgrass} (dotted), and significantly higher than rotation inferred at any depth in the solar interior\cite{beck_schou}.
{\bf b}, Northward meridional flow (diamonds).
The magnetic feature rate\cite{komm_mer} (solid) is again similar.
For comparison, the dashed line shows a time-distance measurement\cite{giles},
while the squares show the anomalous results obtained using correlation
tracking with a 24~hr delay.}
\vspace{10.cm}
\end{figure}

\newpage

\begin{figure}[t]
\centerline{\epsfig{file=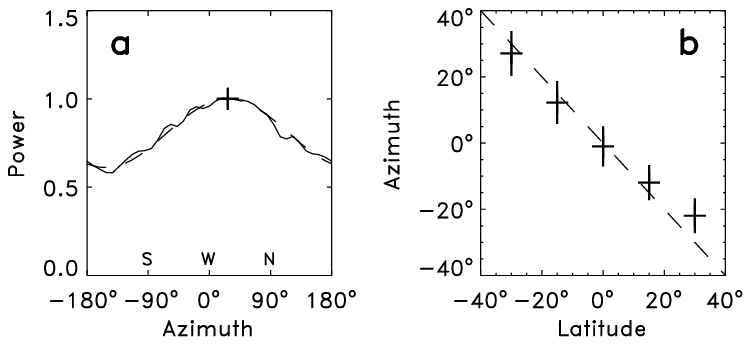, width=15.cm}}
\vspace{1.cm}
{\normalsize {\bf Figure 3} Anisotropic wave power. {\bf a}, Directional distribution of power at latitude $-30^\circ$ (solid line). The azimuth  of maximum power (cross) is measured by fitting a sine (dashed). {\bf b}, Azimuth of maximum power versus latitude, with 1-$\sigma$ error bars. Wave power is maximum in the direction of rotation and toward the equator in both hemispheres. The dashed line is drawn for the azimuth equal to the negative of the latitude.} 
\vspace{10.cm}
\end{figure}

\newpage

\begin{figure}[t]
\centerline{\epsfig{file=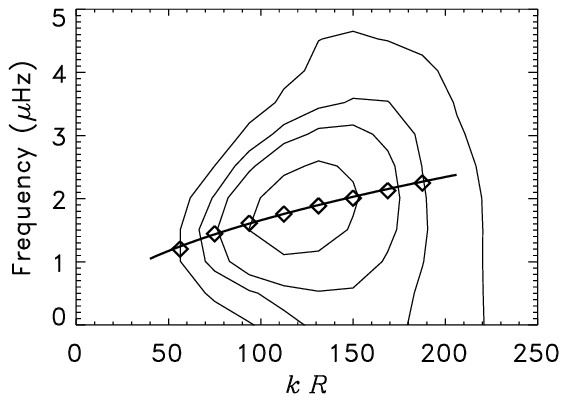, width=15.cm}}
\vspace{1.cm}
{\normalsize {\bf Figure 4} Power spectrum corrected for
rotation and meridional circulation and averaged over azimuth. 
Contour lines are drawn at 15\%, 30\%, 45\%, and 75\% of the maximum value.
The diamonds show the frequencies, $\omega_0(k)$, deduced from 
the fits (see Fig. 1). The empirical dispersion relation $\omega^2 = a k$ with $a=7.5\times 10^{-4}$~\mss is overplotted for reference (thick line). 
We did not find a significant latitudinal variation in the dispersion relation, for latitudes less than 35$^\circ$.   The relatively large spread of the power indicates that the waves have a low quality factor. The width of the power increases rapidly with $k$, presumably due to the increased efficiency of the damping on small scales. Note that the distribution of power as a function of frequency is only affected by the known temporal window function, while the wavenumber dependence includes effects of the telescope optics and of the time-distance analysis.
}
\vspace{10.cm}
\end{figure}

\end{document}